# A color-switchable ring-shaped random laser in momentum space

Yaoxing Bian[1], Xiaoyu Shi[2], Mengnan Hu[1] and Zhaona Wang[1,*]

**Abstract:** A color-switchable random laser is designed through directly coupling random laser with a commercial optical fiber. By using a simple approach of selectively coating the random gain layer on the surface of fiber, the red and yellow random lasers are respectively achieved with low threshold and good emission direction due to the guiding role of optical fibers. Moreover, the unique coupling mechanism leads to the random lasing with ring-shape in momentum space, indicating an excellent illuminating source for high-quality imaging with an extremely low speckle noise. More importantly, random lasing with different colors can be flexible obtained by simply moving the pump position. The results may promote random lasers' practical applications in the fields of sensing, in vivo biologic imaging, high brightness full-field illuminating.

## 1. Introduction

In recent years, lasers have been considered as potential illumination sources for the next generation of projectors[1]. However, the high spatial coherence of lasers results in coherent imaging artifacts caused by interference that occurs during image formation. To improve its' application in the laser display and imaging, some technique schemes have been utilized to passively decrease the spatial coherence of lasing for suppressing coherent artifacts through sacrificing the time[2-5]. Therefore, the active methods of eliminating the coherent noise are highly desired for the next generation of real-time display technology.

As a new type of laser, random lasers utilize multiple scattering to achieve stimulated radiation amplification, avoiding the complex manufacturing process and high cost of conventional resonators[6-12]. Random lasers have becoming a hot research field in the international laser science community due to their special characteristics of low cost, small size and ease of integration, demonstrating versatile application prospects in bio-detection[13, 14], information security[15], and integrated photoelectron[16]. More importantly, the low spatial coherence of random lasing naturally makes up for the shortcomings of traditional lasers in full-field imaging and display. And random lasers are looked as an ideal illumination light source for speckle-free full-field imaging[17-22]. However, the randomness of the emission direction seriously hinders their practical applications[23, 24]. Several technical solutions have been proposed to control the emission direction of a random laser by coupling the random laser with optical fiber or waveguide structure[25-27]. Especially, the plasmonic random lasers were designed on the fiber end facet[28] and in polymer fiber[29] to control the emission direction of the random lasing. These fiber sources are receiving more and more attention due to their directionality, long-distance transmission and easy integration. Notably, the output from these light sources are uniform circular spots in the momentum space. Considering the ring-shaped light sources demonstrating great advantages in suppressing coherent noise[30] and super-resolution imaging[31], it is necessary to design ring-shaped random lasing to further improve the quality of imaging for a wider range of application requirements. In addition, the implementation of color-regulated random lasers is critical to a real illumination source, and researchers have continually explored ways to regulate the frequency range of random lasers from ultraviolet[32], visible[33], to near-infrared[34]. And a color-switchable random laser with highly integration and easy implementation has still been highly desired in the field of illumination.

In this work, a color-switchable random laser is proposed by selectively coupling the random gain medium with the commercial optical fiber. By coating different gain dye films, the red and yellow random lasers with low thresholds of 0.1335 and 0.1193 MW cm$^{-2}$ are achieved, respectively. The obtained random lasing have the unique characteristics of small propagation constant along the optical fiber. Moreover, the color of random lasing can be selectively achieved by simply changing the pumping position. As an illumination source, the random lasing with ring-shape in momentum space assure the high-quality imaging with an extremely low speckle noise. The results may fasten the practical applications of the random laser in speckle-free full-field imaging, laser display, bio-sensing and integrated optics.

## 2. Experimental Methods

The fabrication process of the fiber source is illustrated in **Fig. 1**a. The typical gain materials used in this experiment are: 4-(Dicyanomethylene)-2-tert-butyl-6-(1,1,7,7-tetramethyljulolidin-4-yl-vinyl) -4H-pyran(DCJTB, Tokyi Chemical Industry) and Pyrromethene567 (PM567, Sigma). The fabricated process of the fiber source is simply supplied as follows. First, polydimethylsiloxane (PDMS, *n* = 1.41) solution is mixed with cross-linking solution with the ratio of 1:10. Then, TiO$_2$ nanoparticles are dispersed in the dye-doped acetone solution (DCJTB at 1.5 mg mL$^{-1}$ or PM567 at 1.25 mg mL$^{-1}$) to obtain TiO$_2$ dispersion with a concentration of 0.9 mg mL$^{-1}$. The dye-doped TiO$_2$ dispersion and PDMS are mixed at a volume ratio of 1: 5 in an ultrasonic tank for 15 min and then vacuum them for 40 min to remove the air bubble. Finally, the mixture is dipped onto a clean fiber (TAIHAN Fiber Optics) to flow naturally. The length of the fiber is about 60 mm, the core diameter is 50 μm and the diameter of the cladding is 125 μm. The refractive indexes of the core and cladding layer are 1.54 and 1.52, respectively. The sample is placed in a drying oven at 80 °C for 3 hours to complete the cross-linking polymerization and drying. After cooling to room temperature, a fiber source is realized. Random lasers with different colors can be fabricated by coating different dye polymers at different locations on the surface of the fiber. And the polymer film with different thicknesses can be fabricated by controlling the dipping times.

## 3. Results and Discussion

The optical photo, micrograph and the scanning electron microscope (SEM, Hitachi SU8010) of the obtained samples are shown in Fig. 1b-e, respectively. The polymer film with a thickness of about 1.2 μm is relatively uniformly coated on the fiber surface. Figure 1f shows the used TiO$_2$ nanoparticles with a diameter of around 50 nm. As the scatterer, TiO$_2$ nanoparticles can provide effective feedback for random systems since its refractive index is much larger than the refractive index of PDMS[35, 36].

The experiment setup for the random laser is shown in **Fig. 2**a. The sample (S) is mounted on a stage and optically pumped by a frequency-doubled and Q-switched neodymium doped yttrium aluminum garnet (Nd: YAG) laser with the wavelength of 532 nm, 10 Hz repetition rate, and 8 ns pulse duration passing through a half wave plate (HWP) and a Glan prism (GP) for easily changing the pump power density. Then the pulses is split into two beams by a neutral beam splitter (NBS). One beam is measured by the optical power meter (OPM) for monitoring the pump power density, and the other beam is used to pump the sample through the three mirrors ($M_1$, $M_2$ and $M_3$) for controlling the pumping position of the fiber sample (the inset of Fig. 2a) to achieve the emission of random lasing. The emission spectra are measured by using a spectrometer of Ocean Optics model Maya Pro 2000. All the experiments are performed at room temperature. In our experiments, the detector is approximately 25 mm from the pump position for measuring the lasing spectra unless otherwise specified.

The designing mechanism of the ring-shaped random lasing in momentum space is schematically demonstrated in Fig. 2b. When the DCJTB-doped polymer film is pumped, the light with any directions are generated and scattered by the nanoparticles and the interface in the polymer layer. And random lasing resonance will be built up in the random system as the total gain larger than the loss. Some of the obtained random lasing are attenuated by directly radiating from the gain layer into the air and/or passing through the optical fiber surface into the air, some are coupled into the optical fiber for long-distance transmission and emission from the end surface of the fiber. According to the refraction law and the total reflection condition, the radiation angle $\gamma$ is greater than 39.91° relative to the axis of the fiber when the random lasing is emitted from the two ports of the fiber core as shown in Fig. S1 (Supporting Information). Based on the above analysis, the random lasing is looked as a uniform plane light source at the fiber port, and has a circular spot in the real space in Fig. 2d. When random lasing passing through a lens, the beam is ring-shaped corresponding to the momentum space as shown in Fig. 2e.

**Figure 3**a presents the absorption and photoluminescence spectra of the DCJTB ethanol solution, demonstrating strong absorption for 532 nm pumping pulses and a fluorescence peak at 660 nm. Thus, the DCJTB molecules in the random system can be well excited by the pumping pulses and the emission spectra of the random laser at various pump power densities are shown in Fig. 3b. Only a broad spontaneous emission spectrum centered at 628 nm is observed when the pump power density is less than 0.1591 MW cm$^{-2}$. As the pump power density exceeding 0.1591 MW cm$^{-2}$, several discrete narrow peaks with a linewidth of about 0.6 nm are clearly observed, indicating that coherent resonant feedback is built up in the sample[19, 37]. To calculate the effective cavity length of the random laser, the power Fourier transforms (PFTs) of the spectrum at a power density of 0.9350 MW cm$^{-2}$ (Fig.

3b) is calculated and presented in Fig. 3c. The spatial dimensions can be obtained from the formula $p_m = mnL_c/\pi$, where $p_m$ is a Fourier component, *m* is the order of the Fourier harmonic, *n* represents the refraction index of the gain medium, and $L_c$ is the localized cavity dimension[38, 39]. According to the statistical distribution of the cavity length in the illustration, the effective optical cavity length $L_c$ is calculated to be 122.53 μm ($n = 1.4$). The variation of maximum radiation intensity and full width at half maximum (FWHM) of the random lasing mode of 635 nm with the pump power density are shown in Fig. 3d. When the pump power density exceeds 0.1335 MW cm$^{-2}$, the peak intensity rapidly increases and the line width sharply decreases to sub-nanometer with increasing the pump power density. This significant feature indicates the fabricated random laser with a working threshold of about 0.1335 MW cm$^{-2}$.[40, 41]

The effect of pumping angle $\varphi$ on the performance of random lasing is also carefully studied and shown in Fig. S2 (Supporting Information). Several spikes are generally observed by changing $\varphi$ from 20° to 160° at a pump power density of 0.6565 MW cm$^{-2}$ (Fig. S2b), demonstrating that the multi-mode random lasing can be achieved at an extreme wide angle range. The corresponding threshold remains almost unchanged when the pumping angle $\varphi$ increasing from 20° to 160° as shown in Fig. S2c. Specially, the threshold remains at a small value of 0.1194 MW cm$^{-2}$ when the range of $\varphi$ in [60°,130°]. The stability property of the threshold gives random laser wider implementation conditions.

The effect of thickness *d* of polymer coating on the output performance of fiber-coupled random lasers is also demonstrated as shown in Fig. S3 (Supporting Information) by carefully controlling the dipping times. When the random lasers are vertically pumped by the pulses with a power density of 0.2188 MW cm$^{-2}$, the random lasing intensity increases with *d* rising from 1.2 to 24 μm, following a red-shift behavior of the center wavelength as large as about 10 nm. Larger thickness of the film can provide larger gain for random lasing and a longer period in the gain region which leads to a strong reabsorption of dye molecules and then the spectral red-shift of random lasing[38, 42]. The wavelength shift phenomenon supply an optional approach to regulate the emission wavelength of random laser by controlling the thickness *d* of the coated polymer. In addition, the intensity of random lasing after different distance propagation in optical fiber is detected and shown in Fig. S4 (Supporting Information) by maintaining the pump position and power density unchanged. Coherent random lasers can still be detected with propagation distance *L* increasing to 20 cm, although the intensity decreases with the propagation distance. The result shows the designed fiber source could be used in the integral optics.

The emission directionality of the designed sample is demonstrated in Fig. 4b and S5 (Supporting Information) by keeping the detection distance as 25 mm from the pumping position and changing the detection angle $\alpha$ relative to the fiber axis in Fig. 4a. The relationship between the random lasing intensity and $\alpha$ at a pump power density of 0.9947 MW cm$^{-2}$ is shown in Fig. 4b. A maximal intensity is observed in a very small angular range along the fiber axis which is 33 times as much as that in the 45° direction. This proves that the proposed fiber source has good directionality, which is better than the plasmonic random laser on the optical fiber end facet[28]. Moreover, the spectral stability of the sample is revealed and shown in Fig. 4c-d by measuring the spectrum 100 times at a pump power density of 0.7758 MW cm$^{-2}$. Notably, the intensity of random laser is relatively stable after 100 measurement cycles, indicating that the random laser has good repeatability and spectral stability.

The color of random lasing from the fiber source can be flexibly switched by integrating different color random lasers in one optical fiber as shown in Fig. 5a. And a two-color random laser is fabricated as shown in Fig. S6a (Supporting Information) by using the gain material of Pyrromethene567 (PM567) for yellow and DCJTB for red random laser. Thus, flexible switching of yellow and red random lasing can be achieved by changing the pump position. When the yellow random laser is pumped by 532 nm pulses (Fig. S6b, Supporting Information), the emission spectra are obtained in Fig. 5b at different power densities. Similar to the above-mentioned red random laser, the coherent yellow random lasing are obtained with a FWHM of about 0.5 nm when the pump power density is larger than the threshold of 0.1193 MW cm$^{-2}$ as shown in Fig. 5c. The results demonstrate the potential of the proposed fiber source in achieving multi-color random lasing. Moreover, the fiber source can also be poorly switched between laser and random laser by changing the pumping position on the sample. When pumping the polymer-covered fiber, random laser is achieved in Fig. S7a (Supporting Information) as an example. And a laser of 532 nm is obtained when the pumping position is on the uncoated fiber as shown in Fig. S7b. These results prove the proposed fiber source can be flexibly switched between laser and random laser.

The ability of the fabricated random lasing as an illumination light source of the microscopic imaging in eliminating coherent artefacts is further demonstrated in **Fig. 6**. The optical setup of microscopic imaging is shown in Fig. 6a. The laser is collected by a lens to illuminate the object and/or the scattering film to form the illumination path. An objective lens (100×) is used to collect the object information and a camera is used to record the resultant image for the imaging part. Firstly, the role of preventing speckle formation is directly revealed by taking the images in Fig. 6b and 6c when random lasing and the 633 nm He-Ne laser directly passing through a frosted glass with the coarse surface, respectively. Speckle is obviously observed in Fig. 6c when using the narrowband laser, but the obtained images with random lasing illumination present uniform optical field without any observable speckle in Fig. 6b. To quantitatively analyze the speckle suppression effect, the speckle contrast C is defined as $C = \sigma_I / \langle I \rangle$, where $\sigma_I$ is the standard deviation of the image intensity and $\langle I \rangle$ is the average intensity of the image[17, 22]. The calculated speckle contrasts are $C$=0.41 for the 633 nm laser illumination and $C$=0.02 for random lasing illumination which is lower than that of the polymer fiber random laser[36]. The extremely low speckle contrast attributes in the low spatial coherence and the ring-shape in momentum space of our designed random lasing[30]. Then, the ability of improving image quality of the fabricated random laser is approved by comparing the images of lens paper with random structures illuminated by the two light sources described above. The image based on the narrowband laser exhibits obvious speckle patterns within and near the paper fibers. These artificial intensity modulations corrupt the image as shown in Fig. 6e. However, the random laser can well eliminate the coherent artefacts and produce a clear image of the random object in Fig. 6d. Finally, the role of eliminating speckle of random lasing is demonstrated in a strong scattering environment by introducing a scattering film of frosted glass in front of the lens paper in the illumination path. Under 633 nm laser illumination, high spatial coherence results in strong speckle phenomenon and the image of the lens paper is completely submerged in speckle noise beyond recognition as shown in Fig. 6g. When illuminating with the obtained random laser, speckle noise is precluded due to the low spatial coherence, leading to a uniform background signal and a clean image of the lens paper in Fig. 6f even in a strong scattering environment. These results further prove that the proposed fiber source has a good application prospect in the field of full-field speckle-free imaging and display.

4. Conclusions

A color-switchable fiber source of random laser is fabricated by coating the polymer film on the surface of a commercial optical fiber. The sample possesses novel, unique and complement features in comparison with conventional random laser devices. Firstly, this is a new approach of fabricating the fiber source by coupling of random laser and the commercial optical fiber. The obtained fiber source has the advantages of directional emission, flexibility, easy integration, low special coherence and so on. Moreover, it is very convenient to switch the output characteristic of the fiber source between different color random lasing by mechanically controlling the pumping position. Secondly, the output spots of the random laser from the sample are circular in real space, but ring-shaped in momentum space due to the unique coupling mechanism of the random laser and optical fiber. This unique feature provides a new way to shape the beam for further suppressing coherent noise and achieving super-resolution imaging. Finally, the red and yellow random lasers coupled with the optical fiber have low thresholds of 0.1335 and 0.1193 MW cm$^{-2}$, respectively. The low working threshold assure the sample good stability and repeatability. As a light source, the achieved random laser presents the amazing potentiality in high-quality speckle-free microscopic imaging under different scattering environments. The newly designed fiber source may lead to further practical applications in the fields of sensing, in vivo biologic imaging, high brightness full-field illuminating and interference detection.

## Supporting Information

Supporting Information is available free of charge on the Royal Society of Chemistry Publications website.

## Conflicts of Interest

The authors declare no conflict of interest.

## Acknowledgements

The authors thank the National Natural Science Foundation of China (grant Nos. 11574033, and 11674032), Beijing cooperative construction project, Beijing Higher Education Young Elite Teacher Project and the Fundamental Research Funds for the Central Universities for financial support.

## Notes and references

[1]*Department of Physics, Applied Optics Beijing Area Major Laboratory, Beijing Normal University, Beijing, China, 100875, *Email: zhnwang@bnu.edu.cn*

[2]*College of Applied Sciences, Beijing Engineering Research Center of Precision Measurement Technology and Instruments, Beijing University of Technology, No. 100 Pingleyuan Rd., Beijing, China, 100124*

**Figure Captions:**

**Fig. 1**

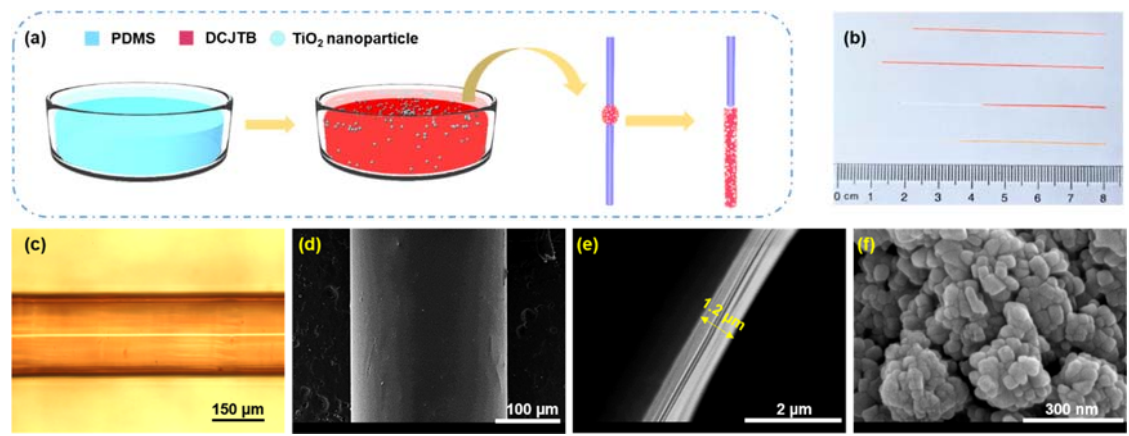

**Fig.1. Preparation and characterization diagram of the fiber source.** (**a**) Specific steps for fabricating a random laser coupled with the optical fiber. (**b-c**) The optical photo and micrograph of the fabricated fiber source. (**d-e**) The scanning electron microscope (SEM) images of the polymer-coated fiber. The side view (**d**) and cross-section view (**e**) are shown. (**f**) The SEM image of $TiO_2$ nanoparticles with a diameter of around 50 nm.

**Fig. 2**

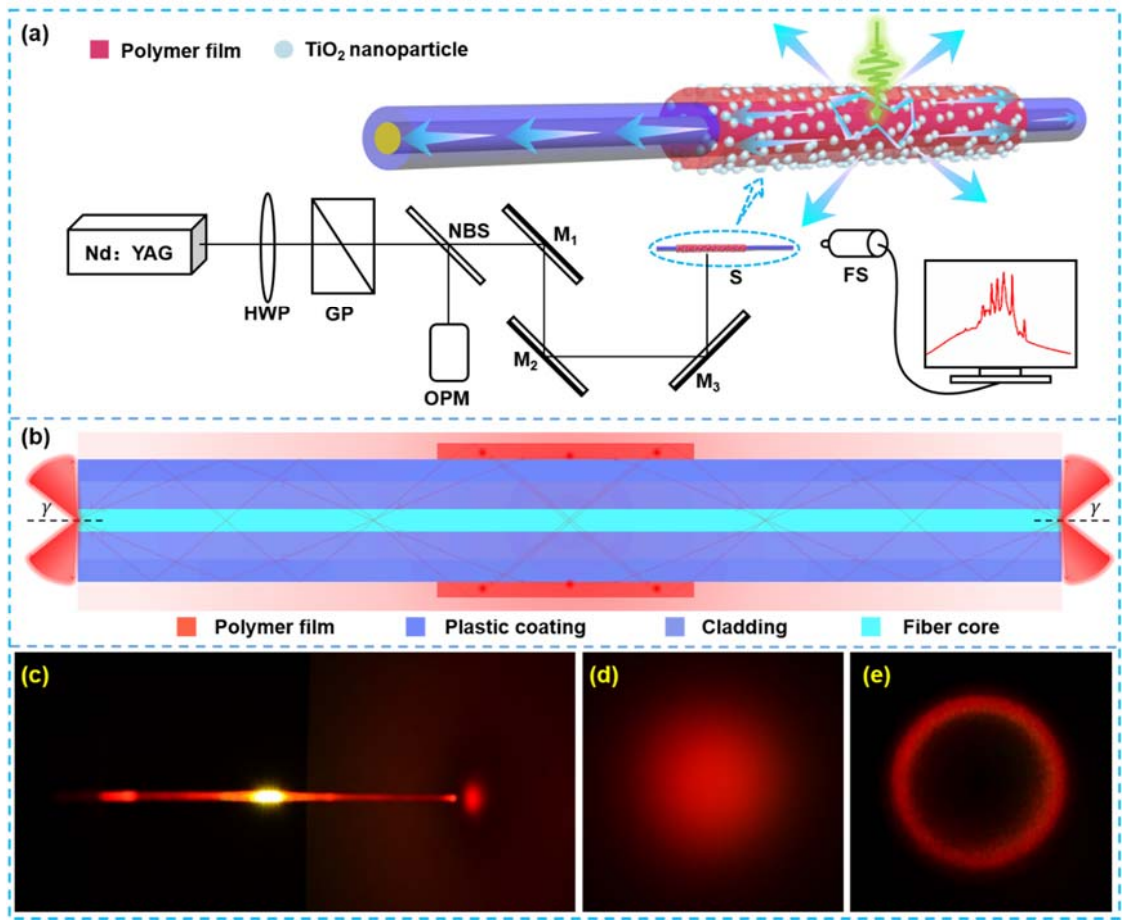

**Fig. 2. Design mechanism of the fiber source.** (**a**) Schematic diagram of the optical path for the fiber sources. Laser: a frequency-doubled and Q-switched neodymium doped yttrium aluminum garnet (Nd: YAG) laser with a wavelength of 532 nm, a pulse duration of 8 ns, a repetition rate of 10 Hz and an diameter of 8 mm; HWP: half wave plate; GP: Gran prism; NBS: neutral beam splitter; OPM: optical power meter; $M_1$, $M_2$ and $M_3$: three mirrors, and $M_3$ is placed on the motorized translation stage; S:the fabricated fiber; FS: fiber spectrometer. (**b**) The schematic diagram of random lasing coupled into an optical fiber. $\gamma$ is the radiation angle of the random lasing emitting from the fiber core. (**c**) Optical photo of the random laser when the DCJTB film is pumped. (**d-e**) The photographs of the random lasing spot in real space (d) and momentum space (e).

**Fig. 3**

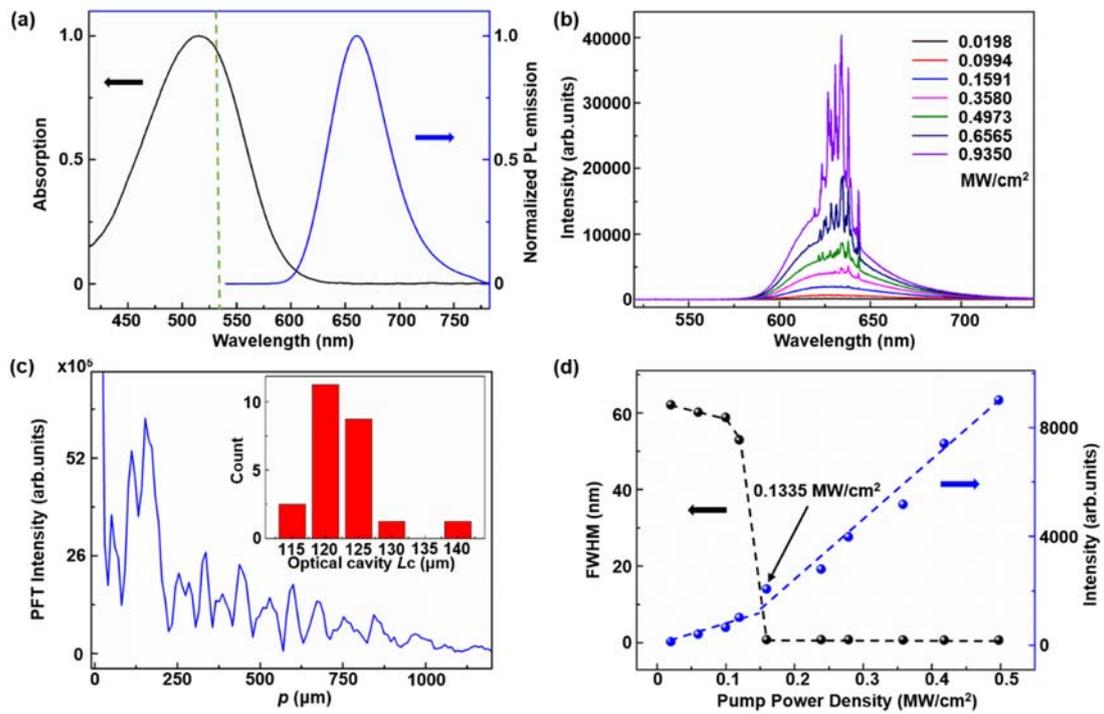

**Fig. 3. Performances of the fabricated red random laser.** (**a**) The absorption and photoluminescence spectra of the DCJTB alcohol solution. (**b**) The emission spectra of the random laser at various pump power densities. (**c**) The power Fourier transform of the red random lasing spectra. The inset is the statistical distribution of the optical cavity length, demonstrating an average cavity length of 122.53 μm. (**d**) The peak intensity and full width at half maximum of the lasing mode centered at 635 nm vary with the pump power density, indicating a threshold of 0.1335 MW cm$^{-2}$.

**Fig. 4**

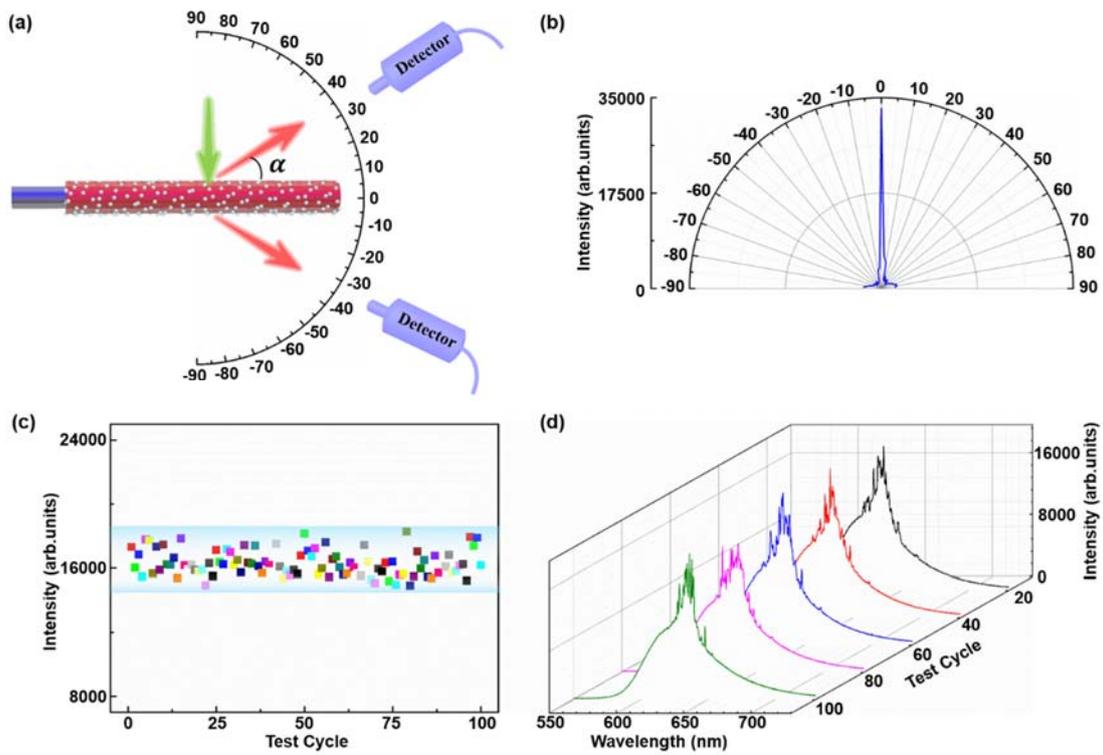

**Fig. 4. Directional emission and stability of the fiber source.** (**a**) A schematic diagram of the intensity measurement at different emission directions when the pump beam is perpendicularly incident to the surface of the polymer film. (**b**) The relationship between the peak intensity and the detection angle $\alpha$ at a pump power density of 0.9947 MW cm$^{-2}$. (**c**) The dynamic laser intensities of 100 times at a pump power density of 0.7758 MW cm$^{-2}$. (**d**) The emission spectra of random laser measured at 20, 40, 60, 80, 100 cycles, respectively.

**Fig. 5**

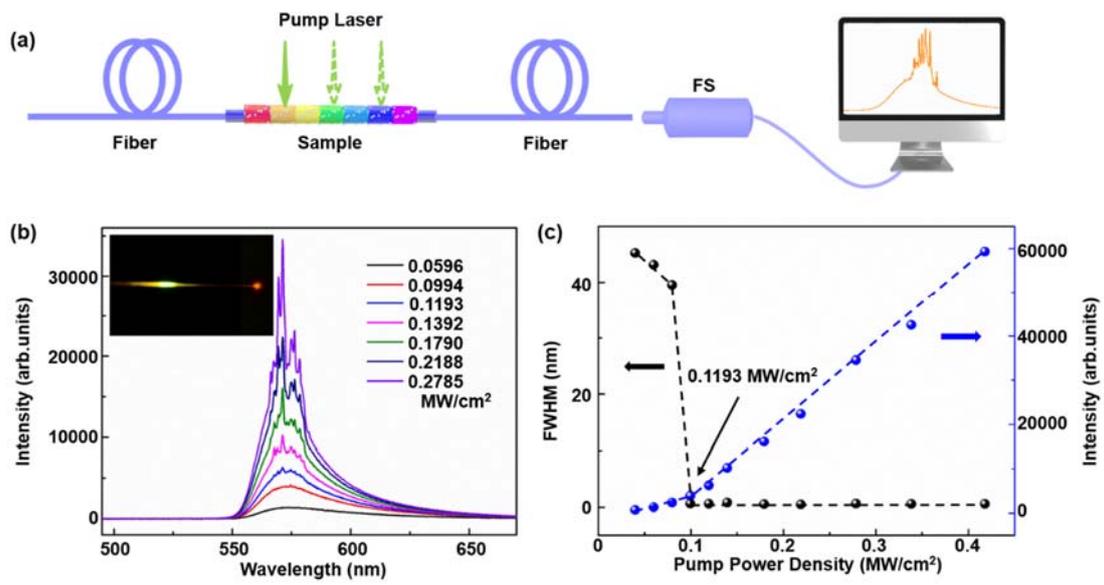

**Fig. 5. Multi-color random laser coupled with optical fiber.** (**a**) The optical path schematic diagram of switchable multi-color random lasers. FS: fiber spectrometer. (**b**) The emission spectra of the yellow random laser at different pump power densities. And the illustration is its optical photo. (**c**) The intensity and FWHM of the yellow random lasing varies with the pump power densities, showing a threshold of around 0.1193 MW cm$^{-2}$.

**Fig. 6**

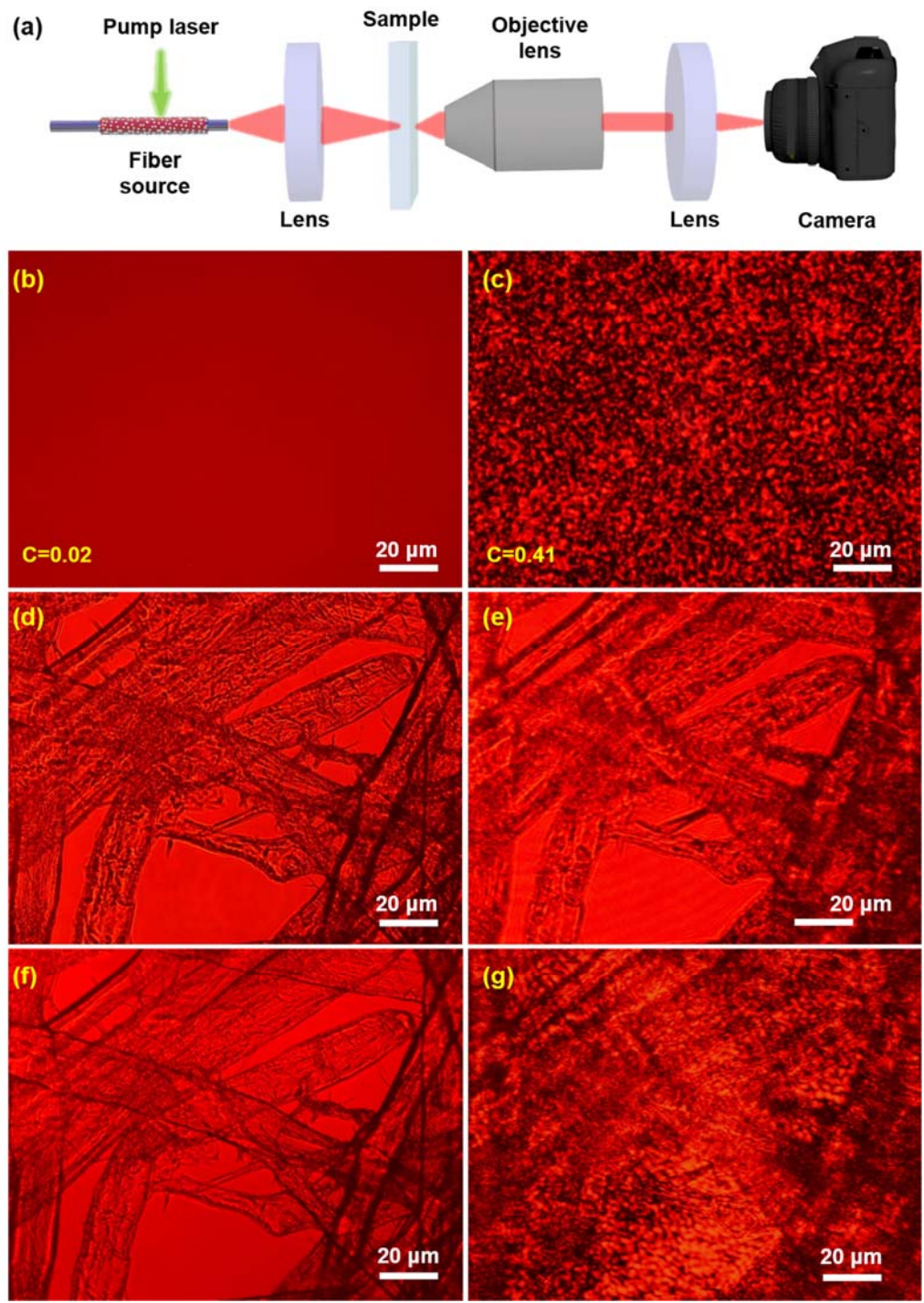

**Fig. 6. Application of random laser in speckle-free full-field imaging.** (**a**) Schematic of the experimental set-up for microscopic imaging. Random laser and He-Ne laser are separately used as the illumination sources. (**b, c**) Optical images of the speckle pattern by using the obtained random laser (b) and a 633 nm laser (c) as the illumination source. The calculated speckle contrasts are shown in the left corner of each image. (**d, e**) Optical images of the lens paper under the illumination of the random lasing (d) and a 633 nm laser (e). (**f, g**) Optical images of the lens cleaning paper under the illumination of the obtained random laser (f) and a 633 nm laser (g) in a strong scattering environment by introducing a frosted glass in the illumination path.

**Table of contents:**

# A color-switchable ring-shaped random laser in momentum space

Yaoxing Bian[1], Xiaoyu Shi[2], Mengnan Hu[1] and Zhaona Wang[1,*]

[1]*Department of Physics, Applied Optics Beijing Area Major Laboratory, Beijing Normal University, Beijing, China, 100875*
[2]*College of Applied Sciences, Beijing Engineering Research Center of Precision Measurement Technology and Instruments, Beijing University of Technology, No. 100 Pingleyuan Rd., Beijing, China, 100124*
*\*Email: [zhnwang@bnu.edu.cn](zhnwang@bnu.edu.cn)*

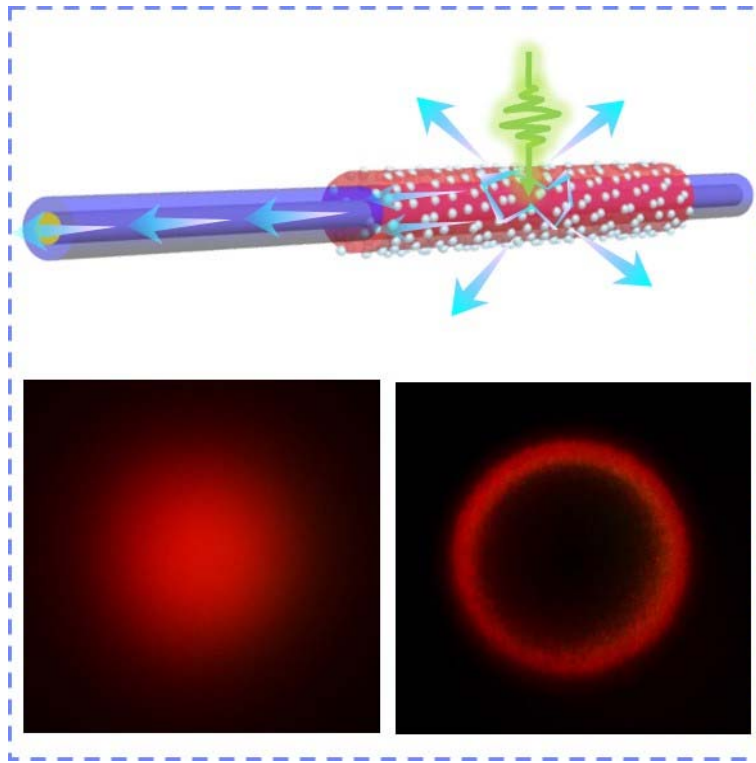

A color-switchable random laser is proposed by coupling the random laser with the commercial optical fiber and mechanically controlling the pump position. The output spots of the random laser are circular in real space but ring-shaped in momentum space. The unique characteristic assure the newly designed random laser wider applications in the fields of in vivo biologic imaging, high brightness full-field illuminating and integrated optics.